\documentclass[times,twocolumn,final]{elsarticle}
\usepackage{ifthen}
\setboolean{preprint}{true}

\usepackage{ifdraft}

\ifthenelse{\boolean{preprint}}{
\newcommand{\snm}[1]{#1}
\usepackage{cuted}
\usepackage{hyperref}
\usepackage{geometry}
\usepackage{multirow}
\usepackage{color}
 \geometry{twoside,
  paperwidth=210mm,
  paperheight=280mm,
  textheight=635.35pt,
  textwidth=467.75pt,
  inner=32pt,
  outer=43pt,
  top=65pt,
  bottom=40pt,
  headheight=12pt,
  headsep=12pt,
  footskip=16pt,
  footnotesep=28pt plus 2pt minus 6pt,
  columnsep=18pt
 }
 \definecolor{gray10}{gray}{.9}
}{\usepackage{medima}}

\usepackage{framed,multirow}

\usepackage{amssymb}
\usepackage{latexsym}

\usepackage{url}
\usepackage{xcolor}

\usepackage{hyperref}

\definecolor{newcolor}{rgb}{.8,.349,.1}

\ifthenelse{\boolean{preprint}}{\journal{ }}{\journal{Medical Image Analysis}}

\usepackage{siunitx}
\usepackage{textgreek}
\usepackage{subcaption}
\usepackage{booktabs}
\usepackage{makecell}

\begin{document}

\ifthenelse{\boolean{preprint}}{}{\verso{Balluet Maël \textit{et~al.}}}
\ifthenelse{\boolean{preprint}}{\begin{strip}}{}
\begin{frontmatter}

\title{Neural network fast-classifies biological images using features selected after their random-forests-importance to power smart microscopy.}

\author[1,2]{Maël \snm{Balluet}}
\ifthenelse{\boolean{preprint}}{
\author[1,6]{Florian \snm{Sizaire}}
}{
\author[1]{Florian \snm{Sizaire}\fnref{fn1}}
\fntext[fn1]{Present address: Biologics Research, Sanofi R\&D, F94400 Vitry sur Seine, France}
}
\author[1]{Youssef \snm{El Habouz}}
\author[3,5,7]{Thomas \snm{Walter}}
\author[2]{Jérémy \snm{Pont}}
\author[2]{Baptiste \snm{Giroux}}
\author[2]{Otmane \snm{Bouchareb}}
\author[1,4]{Marc \snm{Tramier}}
\author[1]{Jacques \snm{Pecreaux}\corref{cor1}}
\ead{jacques.pecreaux@univ-rennes1.fr}
\cortext[cor1]{Corresponding author:Jacques Pecreaux}

\address[1]{CNRS, Univ Rennes, IGDR - UMR 6290, F-35043 Rennes, France}
\address[2]{Inscoper SAS, F-35510 Cesson-Sévigné, France}
\address[3]{Centre for Computational Biology (CBIO), MINES ParisTech, PSL University, F-75272 Paris, France}
\address[5]{Institut Curie, F-75248 Paris, France}
\address[7]{INSERM, U900, F-75248 Paris, France}
\address[4]{Univ Rennes, BIOSIT, UMS CNRS 3480, US INSERM 018, F-35000 Rennes, France}
\ifthenelse{\boolean{preprint}}{
\address[6]{Present address: Biologics Research, Sanofi R\&D, F94400 Vitry sur Seine, France}
}{}
\ifthenelse{\boolean{preprint}}{}{
\received{1 May 2013}
\finalform{10 May 2013}
\accepted{13 May 2013}
\availableonline{15 May 2013}
\communicated{XYZ}}

\sisetup{separate-uncertainty,multi-part-units = single, range-units = single, list-units = single}

\begin{abstract}
Artificial intelligence is nowadays used for cell detection and classification in optical microscopy, during post-acquisition analysis. The microscopes are now fully automated and next expected to be smart, to make acquisition decisions based on the images. It calls for analysing them on the fly. Biology further imposes training on a reduced dataset due to cost and time to prepare the samples and have the datasets annotated by experts. We propose here a real-time image processing, compliant with these specifications by balancing accurate detection and execution performance. We characterised the images using a generic, high-dimensional feature extractor. We then classified the images using machine learning for the sake of understanding the contribution of each feature in decision and execution time. We found that the non-linear-classifier random forests outperformed Fisher’s linear discriminant. More importantly, the most discriminant and time-consuming features could be excluded without any significant loss in accuracy, offering a substantial gain in execution time. It suggests a feature-group redundancy likely related to the biology of the observed cells. We offer a method to select fast and discriminant features. In our assay, a \SI{79.6 \pm 2.4}{\percent} accurate classification of a cell took \SI{68.7 \pm 3.5}{\milli\second} (mean $\pm$ SD, 5-fold cross-validation nested in 10 bootstrap repeats), corresponding to 14 cells per second, dispatched into 8  phases of the cell cycle using 12 feature-groups and operating a consumer market ARM-based embedded system. Interestingly, a simple neural network offered similar performances paving the way to faster training and classification, using parallel execution on a general-purpose graphic processing unit. Finally, this strategy is also usable for deep neural networks paving the way to optimising these algorithms for smart microscopy.

\end{abstract}

\ifthenelse{\boolean{preprint}}{}{
}
\end{frontmatter}
\ifthenelse{\boolean{preprint}}{\end{strip}}{}

	\section{Introduction}
	
	
	The optical microscope, after centuries as an advanced optical device, underwent significant evolutions during the last decades to become the motorised system now controlled by electronic signals. Its variegated modalities make it an unparalleled tool to investigate the living \citep{nketia17}. Beyond academic research, it can automatically image samples in large series, together with the appropriate robots, paving the way to live-cell high content screening (HCS) based on phenotypes \citep{esner18,peng08,sbalzarini16,chen18}. However, the analysis of this data flood is performed posteriorly to the acquisition, limiting the information extracted \citep{singh14}. A smart microscope, able to modify the imaging strategy in real-time by analysing images on the fly, is required to increase the number of images interesting for the biological question (so-called qualified images)\citep{scherf_smart_2015}. By autonomously acquiring rare objects and elusive events, it will not only ease basic-research imaging by saving fastidious searching and waiting for a cell of interest at the right stage but also increase the content of interest in HCS by selecting qualified images, up to become a standard tool of precision medicine similarly to next-generation sequencing \citep{hamilton14,leopold18,klonoff15,djuric17}. The current systems that perform imaging and analysis in tandem alternate acquiring images and analysing them \citep{conrad11,tischer14}. We recently achieved efficient microscope driving \citep{sizaire20,roul15} and here investigate how to perform the real-time object's classification to feedback to it.
	 
	Searching rare and brief events is a booming field beyond the sole microscopy. They often carry significant information about normal or abnormal processes in a broad range of applications \citep{ali15,kaushal18}. Radiologists use such algorithms to assist the medical-doctor diagnosis interactively, calling for reduced image processing delay \mbox{\citep{chartrand17}}. Along a line more demanding of real-time processing, video can be processed to recognise the human activities, in particular, risky or abnormal situations like intrusions or dangerous behaviours \citep{bobick01,zhang17,sargano17}. Similarly, it can support  detecting and diagnosing faults in construction or process industries \citep{koch15,duchesne12}. These situations may result in costly damages, human injuries and require rapid detecting through real-time analysis. We here used a similar approach to detect rare and transient events in living biological samples. 
	
	Very archetypal to these events is the anaphase of cell division when the sister chromatids are separated to be equally distributed to each daughter cell. In human cells, it lasts a few minutes or less in contrast with a cycle of 15 to 30 hours (the repetition time of mitosis) \citep{moran10}. Cell division has received strong attention in fundamental research as its mechanisms are only partially known, as well as in applied research in particular to develop cancer therapies \citep{manchado12,florian16,mcintosh17,rieder03,ciresan13}. Indeed, the spindle assembly checkpoint (SAC) secures the transition to anaphase by ensuring a correct attachment of the chromosomes, essential to their equal partitioning to daughter cells. However, this checkpoint may fail to detect errors or slip, paving the way to cancer \citep{potapova17,sivakumar15}. Unfortunately, the current techniques to investigate these phenomena are invasive, as blocking cultured (human) cells for a few hours at the entry in mitosis by drugs similar to antimitotic ones used in cancer therapies \citep{banfalvi17}. Doing so lets most of the cells reach the threshold of mitosis before the experimenter releases the block to observe all cells undergoing mitosis in a synchronised fashion. Although instrumental, this technique is perturbative, and we propose here to leap towards superseding it by detecting mitosis when they occur rather than triggering them artificially. Along an applied line, targeting mitosis is a cornerstone to designing drugs used in chemotherapy \citep{manchado12}. It implies the ability to fast screen across a library of compounds and quickly assess defect in mitosis and particularly deadlocked mitosis due to unsatisfied SAC \citep{mcintosh17}. Along a medical line, detecting mitosis in patient tissues is classically used for diagnosis as in breast cancer  \citep{wang_mitosis_2014,  hamidinekoo_deep_2018, veta15}. Overall, it makes the automated detection of early anaphasic cells a highly relevant application case.

	Beyond these applications, both fundamental and applied cell microscopy would need an approach to detect rare and short events to instruct the microscope some specific acquisition conditions. Such a system should exhibit three main specifications: perform fast enough to achieve real-time detection; being adaptive to a wide variety of problems (cell types, labellings or events of interest, e.g.) without re-programming or re-optimising; achieve this adapting (training) over a reduced exemplar dataset. While some dedicated image processings allow post-processing of the data and identification of the hits in high content screening \citep{wollmann17,fillbrunn17, mcquin18}, each application resulted from a dedicated development. Furthermore, suitable performance often requires a detailed and long optimisation of the specific program. In particular, algorithms were developed to classify mitotic cells in distinct stages, along time and in live samples \citep{harder_automatic_2009,held_cellcognition_2010,conrad11}. These classifiers may, however, turn to be too slow for real-time since we aimed to acquire and classify images on the fly concurrently. Furthermore, these algorithms are specialised to a given biological situation while we aim at developing a single software adapted to a broad range of applications, i.e. generic. These latter approaches had used to result in poor classification as they involved one or a few generic features \citep{sbalzarini16}. In the last decades, the emergence of machine learning has been a real game-changer and allowed both generic and accurate analysis, and paved the way to new experiments \citep{moen19,nagao20,singh14,sommer13,sbalzarini16,nketia17}. Along that line, we here used a wide variety of features found in the library WND-CHARM \citep{orlov_wnd-charm:_2008}. Key to perform accurate and fast detecting was to select a subset of these features and combine them into an efficient discriminator. It enabled to optimise the code once and for all, without editing it again. The specifics of the application were encoded into a statistical model. Machine learning approaches addressed this need and could be trained easily to each application through a numerical optimising onto a set of labelled images. In contrast to deep learning, it enabled to identify important features and even manually manipulate their selected subset to improve execution time. We then embedded this classifier and adapted it to the case through its training to ensure real-time execution, paving the way to the autonomous microscope \citep{balluet20}. In this article, we proposed a strategy to optimise the selection of features of interest under the constraint of both accurate classification and fast performing. It implies to select features both quick to execute  and discriminant. Amazingly, we found that highly discriminant features could be excluded, provided enough other features were available, without any loss in classification accuracy and with a strong gain in execution time on an ARM embedded system.
	
	\section{Materials and methods}	
	\subsection{Image database}
	\label{imDB}
\begin{figure}[h!]
		\centering
		\includegraphics{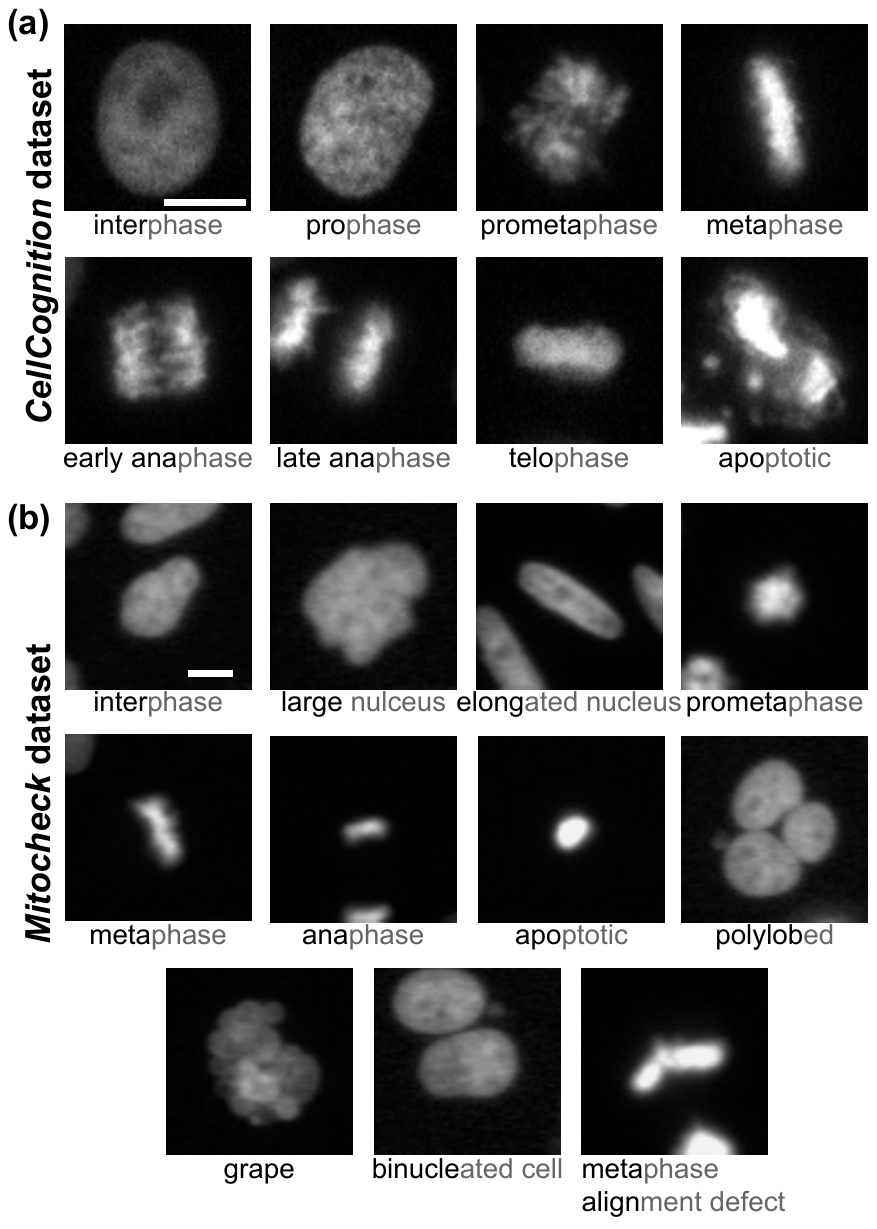}
	
		\caption{\textbf{Datasets used during numerical experiments.} \textbf{(a)} Exemplar vignettes upon $71 \times 71$ pixels cropping images from the CellCognition database. \textbf{(b)} Exemplar vignettes similarly cropped and extracted from the mitocheck database. Class names were abbreviated and written in black font, while the full name appears in grey. They correspond either to cell division phases or specific defects: cells whose nucleus display an elongated, polylobed or grapefruit-like shape, and nuclei reminiscent of apoptotic cells, binucleated ones (usually following a cytokinesis defect) or cells having an issue in aligning the chromosomes during metaphase, usually due to lagging chromosomes or multipolar spindles. A scale bar indicates \SI{10}{\micro\meter} in the first frame, and all vignettes within a dataset are on the same scale.}
		\label{cell_classes}
	\end{figure}

	We built a first image database (termed \textit{CellCognition}) from the CellCognition \citep{held_cellcognition_2010} software demonstration images. It is composed of wide-field fluorescence time-lapses of human Hela Kyoto cells, expressing histone H2B and \textalpha-tubulin markers, which revealed the chromosomes and the microtubules respectively. Images are acquired at three different positions with a 20x dry objective and taken with a time interval of \SI{4.6}{\minute}. Each field contained 206 images of $1392 \times 1040$ pixels, including multiple cells. The corresponding annotations classified the cells between 8 classes, including the six mitotic phases and indicated the centre of the object\citep{held_cellcognition_2010}. We built a database of  $71 \times 71$ pixels vignettes corresponding to classified cells extracted from the fields. Cells exemplary of each class are presented in Fig. \ref{cell_classes}a. We removed multiple instances of the same cell appearing at different stages and thus in distinct classes. We also discarded randomly chosen vignettes to equilibrate the dataset. We obtained 159 vignettes altogether, specifically 20 per class, except apoptosis showing 19 vignettes. This low number of cells was in line with our application in cell biology since large training sets are not achievable for experimental reasons.
	
	To demonstrate that our classification method is generic, we used a second database, termed \textit{mitocheck}. It is based on the class definitions published in \citep{neumann10}. With respect to this paper, we significantly increased the number of samples in each class. In addition, we added a second artefact class "Focus". For annotation, we preselected experiments that showed phenotypes according to the analysis in \citep{neumann10}, and we manually annotated individual nuclei in these movies without looking at the initial classification. For the dynamic phenotypes, such as prometaphase and metaphase, we sometimes used the time information to decide, in accordance to the procedure in \citep{neumann10}. In total, we annotated 5151 nuclei. It was composed of wide-field fluorescence time-lapses of Hela Kyoto cells, expressing chromatin GFP marker but no \textalpha-tubulin, acquired with a 10x dry objective on Olympus ScanR. Several mitotic phases and defect phenotypes were observed. After equilibration, we obtained  1100 vignettes of $64 \times 64$ pixels dispatched up into 11 classes (100 per class)  (see Fig. \ref{cell_classes}b).

	\subsection{Features extraction}
		
\begin{figure}[h!]
			\centering
				\includegraphics{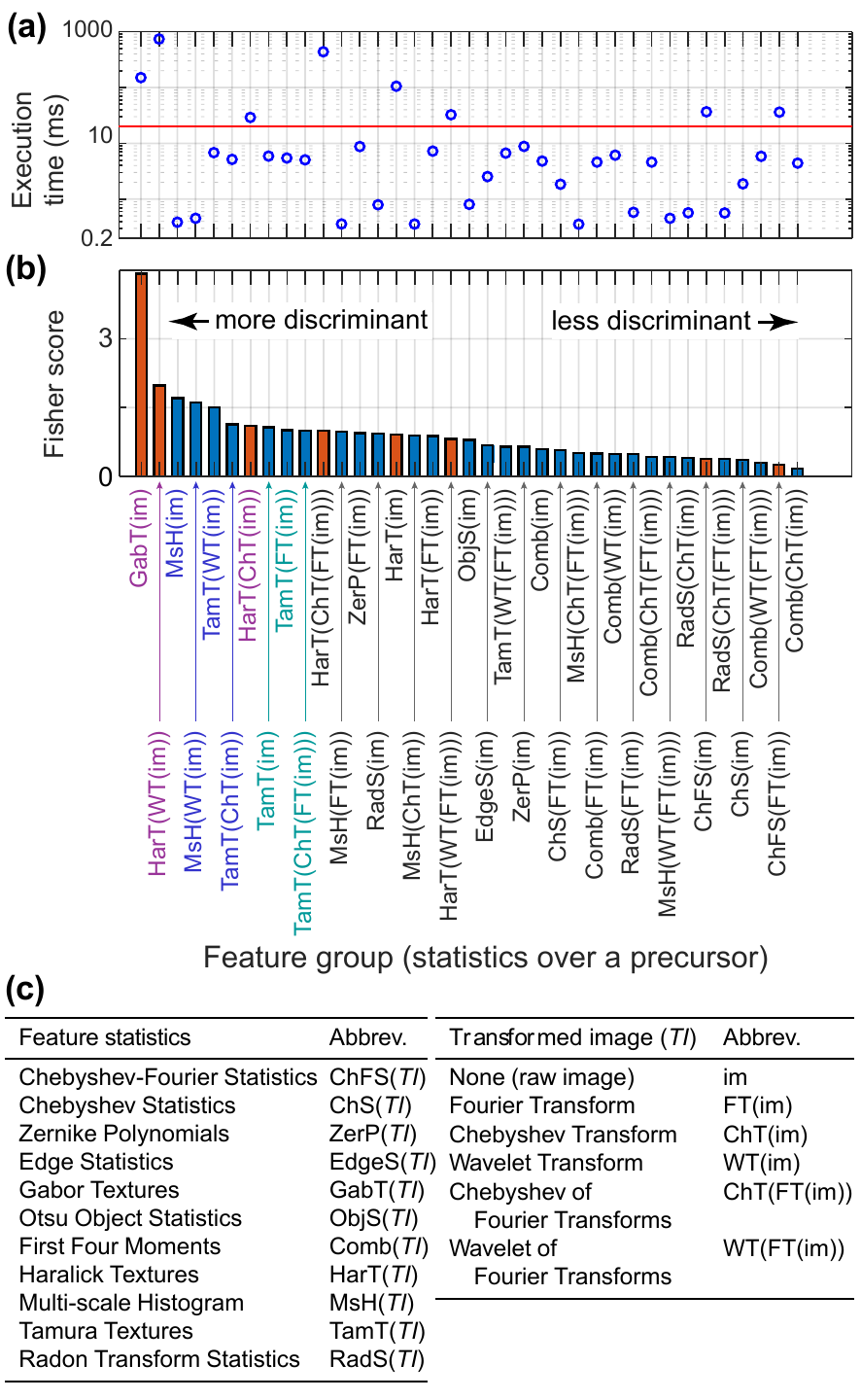}			
			\caption{\textbf{Feature-groups execution time and Fisher's score.} \textbf{(a)} Execution time summed up over feature groups, estimated on an NVIDIA Jetson AGX Xavier embedded system, and \textbf{(b)} the corresponding Fisher's score averaged over the same feature groups (see Methods, §\ref{time_estimate} and §\ref{score_estimate}). \textbf{(c)} (left) Depicts the feature groups by statistics, computed over (right) various feature precursors, i.e. the raw image or its transform. Red bars highlight the feature groups displaying an execution time greater than \SI{20}{\milli\second}. A red line depicts this threshold time in panel (a). Feature-group labels written with colour depict the ones kept for assay using Fisher's linear discriminant (see §\ref{fisher}), specifically the purple and dark blue when considering all feature-groups and the dark and light blue when excluding computationally intensive groups. . When excluding computationally intensive features, the blue ones are also used to complement to 7 groups. CellCognition dataset was used (see Methods §\ref{imDB}).}
			\label{fisher_scores_and_exec_time}
		\end{figure}

		WND-CHARM is a multi-purpose image classifier developed in C++, generating a high-dimension features-vector and using Weighted Neighbour Distances for classification \citep{orlov_wnd-charm:_2008}. We used it to extract edges and objects statistics, multi-scale histograms, four first moments on images subdivision, polynomial decompositions (Chebyshev, Chebyshev-Fourier and Zernike), texture information (Haralick, Tamura and Gabor textures) and Radon transform statistics. In a first step, a transform like Fourier or wavelet could be applied to the raw vignettes to produce a so-called feature precursor, which is an image  (Fig. \ref{fisher_scores_and_exec_time}c, right), on which statistics are extracted (Fig. \ref{fisher_scores_and_exec_time}c, left). Technically speaking, we gather in these statistics some computations that could involve the image (Otsu thresholding for Otsu object statistics case, e.g.) before computing scalar values as statistics (the bright segmented region area in Otsu statistics, e.g.). All features were scalar and were gathered in a 1025-valued vector. Importantly, we performed some optimisation of the WND-CHARM library to reduce its execution time.

	\subsection{Estimating the computing time of features extraction}
	\label{time_estimate}
	To estimate the computing time of a single WND-CHARM feature, we computed it over the single-cell vignettes obtained for instance from CellCognition database, running on an NVIDIA Jetson AGX Xavier embedded system. We then averaged the results over the vignettes of all the dataset.
In particular, we ensured that the execution was sequential on the CPU of the embedded system, without using parallelism. When it comes to estimating the computing time of multiple features, we noticed that the features were not independent. Indeed, for a given group of features, they all correspond to statistics computed on the same \textit{feature precursor}. This latter was either the raw image or a transform computed from it. Several image-transforms could be composed together successively (Fig. \ref{fisher_scores_and_exec_time}c). Notably, the major part of computing time was spent in getting such feature precursors. We thus considered that features were computed by group deriving from the same-precursor. We thus summed up the execution times of all of them within a group, to get the group execution-time.  For instance, in the case of the features based on the Haralick texture, the feature-precursor computation took 90\% to 99\% of the whole computing time (Fig. \ref{fisher_scores_and_exec_time}a).

	\subsection{Estimating the fisher score of features and feature-groups}
	\label{score_estimate}
	The contribution of a feature to the classification was estimated using Fisher's score \citep{orlov_wnd-charm:_2008, bishop_pattern_2006}. For the feature groups as defined above (see §\ref{time_estimate}), we averaged the score of the features over the whole group. Because various statistics within a group might display different scores, such an averaging strategy will favour groups with a majority of well-discriminant features. 
	
	\section{Results}
	\subsection{Classifying based on a single feature was not accurate enough.}
	We set to automatise the microscope by processing images on-the-fly and feeding back to the microcontroller that drove the microscope and its attached devices. To ensure real-time processing, we embedded the processing on a microcontroller as it was designed to execute only one or a few dedicated functions, with real-time constraints, by opposition to a general-purpose computer. It is widely used in fields requiring real-time applications and machine learning algorithms are now available on these platforms. To support the development, we classified mitotic images within 8 classes using the CellCognition example set  \citep{held_cellcognition_2010,CellCognitionImDB} (see Methods §\ref{imDB}) and in particular detected the transition from metaphase to anaphase. We reckoned that the choice of the features could be essential for performance and precision. Therefore, we used the WND-CHARM framework that encompassed a large variety of features \citep{orlov_wnd-charm:_2008}. First, aiming at fast processing, we asked whether a single feature could be sufficient. We computed Fisher's score of each feature (see Methods §\ref{score_estimate}) and found that the most discriminant one  was the area of the segmented image with an Otsu static threshold \citep{otsu79}. The area of Otsu object was highly efficient to discriminate interphase from mitosis. However, this feature was unable to correctly detect anaphase onset since it was mostly sensitive to the surface of the bright objects (Fig. \ref{confMatOtsu}) It called for a multi-feature approach.

	\subsection{Selecting an optimal set of feature-groups using Fisher's linear discriminant.}
	\label{fisher}
				Computing all the features offered by the WND-CHARM library for a $71\times71$ vignette on the ARM microcontroller, was too computationally intensive for several features (Fig. \ref{fisher_scores_and_exec_time}a), thus incompatible with real-time analysis. We foresaw that a small number of features could be combined into a discriminant score, sufficient to discriminate the different mitotic stages. To do so, we opted for a machine learning approach, to help to delineate important features, rather than a deep learning approach. Such an \textit{a priori} choice appeared the most fitted to our lack of large training set and need for fast computation. Indeed, deep-learning-network convolutional layers are computationally intensive, and while optimisation strategies are available for embedded instances like pruning or quantisation \citep{jacob18,molchanov16}, it requires a large training set. We first opted for a linear machine-learning algorithm, specifically the Fisher's linear discriminant \citep{fisher36,duda73}. Indeed such a kernel method, because linear, promised short execution times and was successful in similar problems \citep{muller01,belhumeur97,liyang05,chiang00}. 

\begin{figure}[h!]
			\centering
			\includegraphics{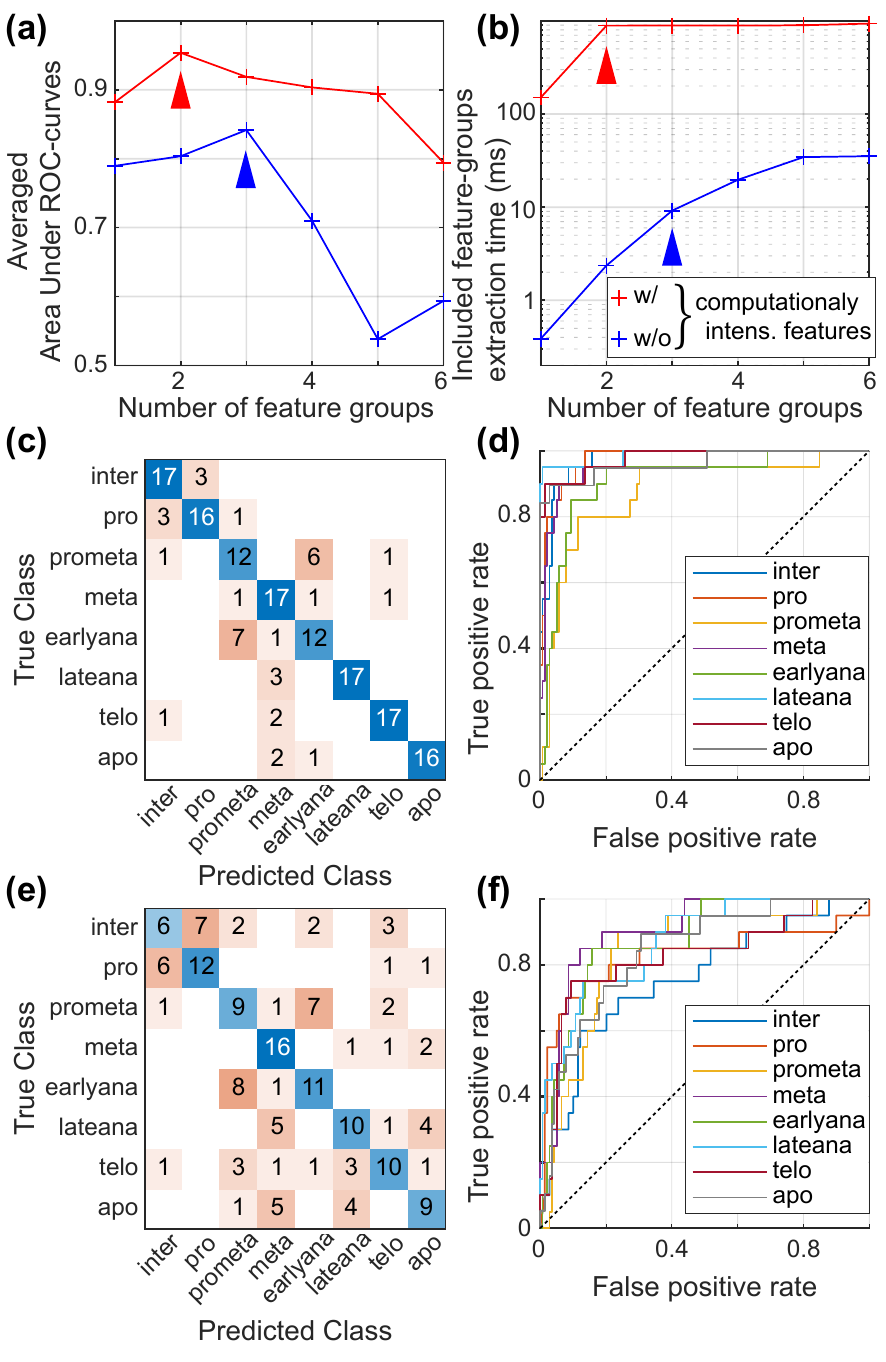}
			
			\caption{\textbf{Classification using Fisher's linear discriminant.} \textbf{(a)} Area Under Curve (AUC) averaged over the classes and \textbf{(b)} execution time for extracting the feature-groups included in the classification, both versus the number of feature-groups used in classification, including (red curve) all available features or (blue curve) only groups with an execution time below \SI{20}{\milli\second} (not computationally intensive). Arrowheads of the corresponding colour depict their optimal number (see §\ref{Fisheroptim}). \textbf{(c)} and \textbf{(e)} report the corresponding confusion matrix for these two-groups (Gabor textures and Haralick over wavelet transform ones), and three-groups (multi-scale histograms over raw vignettes, multi-scale histograms over wavelet transform, and Tamura textures over wavelet transform) optimal cases, respectively, and \textbf{(d)} and \textbf{(f)} are the corresponding ROC curves. Class names are abbreviated after Fig. \ref{cell_classes}a. We used the 5-fold cross-validation over the CellCognition dataset (see Methods §\ref{imDB}).}
			\label{fisher_classification_results}
		\end{figure}

		\label{Fisheroptim}
		We tested Fisher's linear-discriminant classification using the CellCognition dataset (see Methods §\ref{imDB}), in particular, 80\% of the vignettes for training and 20\% for testing through a \textit{k}-fold cross-validation process ($k = 5$). We ranked the feature groups by decreasing Fisher's score (see Methods \ref{score_estimate}). To avoid overfitting, we limited the number of features considered to less than the number of training images. We included the feature-groups in descending fisher score up to that limit. It led to the 7 feature-groups (named in purple and dark blue Fig. \ref{fisher_scores_and_exec_time}b) \citep{shaikhina15,kourou14,foster14}. To find an optimal number of features, we further pruned the feature groups by removing the least discriminant one iteratively until it harmed the overall classification. In further details, we assessed the quality of the classification through the area under the ROC curves (AUC) averaged over the eight classes of our dataset, a classical metric in machine learning \citep{fawcett_introduction_2006}. We measured the maximum AUC when removing the groups and conserved as many groups as needed so that the AUC is not decreased by more than 0.005 from this maximum. It could be achieved without re-training, taking advantage of the linearity (Fig. \ref{fisher_classification_results}a). Such a reduction of the feature-groups number, beyond performance consideration, is essential to cope with the scarcity of labelled images, a commonplace in microscopy for biology and medicine. We obtained the best classification by considering only 2 groups, namely Gabor textures and Haralick calculated from Wavelet transform ones (Fig. \ref{fisher_scores_and_exec_time}b, Fig. \ref{fisher_classification_results}a, red curve and arrowhead). While the classification could be satisfactory with a global accuracy of 78.0\% (Fig. \ref{fisher_classification_results}c and \ref{fisher_classification_results}d), the execution time, \SI{890}{\milli\second}, was incompatible with the on-the-fly classification (Fig.  \ref{fisher_classification_results}b).

		We noticed that the most discriminant feature-groups displayed a score neatly larger compared to the others (Fig. \ref{fisher_scores_and_exec_time}b). However, the two most discriminant groups used for optimal classification were too computationally intensive for our application. We reckoned that they could be removed, keeping a reasonable classification accuracy. In a broader take, we censored all the feature groups, which required more than \SI{20}{\milli\second} to be computed (Fig. \ref{fisher_scores_and_exec_time}a, red line). We again considered 7 feature-groups only to prevent overfitting (named in ligh and dark blue Fig. \ref{fisher_scores_and_exec_time}b). We then selected a subset of the groups, by excluding the least discriminant ones, as explained above. We obtained the best classification using 3 feature groups (Fig. \ref{fisher_classification_results}a, blue curve):  multi-scale histograms calculated from raw vignettes, multi-scale histograms from Wavelet transform of the vignettes and Tamura textures from Wavelet transform. However, while the transition from metaphase to anaphase was still correctly detected, the confusion matrix and the ROC curves, on early and late mitotic phases, showed a clear degradation of the classification (compare Fig. \ref{fisher_classification_results}ef with Fig. \ref{fisher_classification_results}cd). Overall, the accuracy read 52.2\% and class-averaged AUC 0.842 for the three-groups case, compared to 78\% and 0.954, respectively, for the two-groups case including the computationally-intensive features. Using three non-computationally-intensive feature-groups only partially compensated the lack of the two most-discriminant groups and resulted in classification so inaccurate that it could not fit our applicative needs. However and importantly, the feature extraction took only \SI{9}{\milli\second} in the three-groups case, compared to \SI{890}{\milli\second} in the two-groups one, in line with embedded on-the-fly processing.

		Overall, using multiple feature-groups in classification needed a tedious balance between accuracy and execution time, unworkable by a linear machine learning approach. However, we observed a partial redundancy of the features in distinct groups. Importantly, we noticed that the classifying itself took a negligible time, provided that the features were already computed. It called for using non linear classification method to combine the features at the expense of computing time.

		\subsection{Revealing the feature-groups redundancy using random forests.}
		
		We pursued searching for a feature-group subset, fast enough to be used in our real-time application and using a non-linear classifier. We set to use a decision-tree based method as it copes well with the large number of features coupled to the reduced training dataset. We specifically chose the random forests algorithm  \citep{tuv_feature_2009,breiman_random_2001}. It is a machine learning algorithm based on an ensemble of decision trees, that furthermore internally selects the most discriminant features, in line with our goal to use a subset of feature groups. Compared to other non-linear methods, random forests, by this selection process, better avoids over-fitting problems. Practically, we trained 300 decision trees using curvature test to select the best split predictor \citep{loh_split_1997}, and we validated this model using \textit{k}-fold cross-validation with $k = 5$. We empirically determined the number of trees, measuring that more than 300 trees would not improve the classification accuracy (Fig. \ref{RFnbTree}). We first performed the classification using all the 1025 features, and the algorithm training converged. The global accuracy read 81.8\% and AUC  0.974, which is slightly better compared to Fisher's linear discriminant. All the classes were recovered at least as accurately or better by the random forests (Fig. \ref{random_forest_full_dataset_result_AUC_ROC}). This encouraging result confirmed the suitability of the random forests to our problem. However, extracting all the features from the image remained too computationally intensive for our application. 
				
		\begin{figure}[h!]
			\centering
			\includegraphics{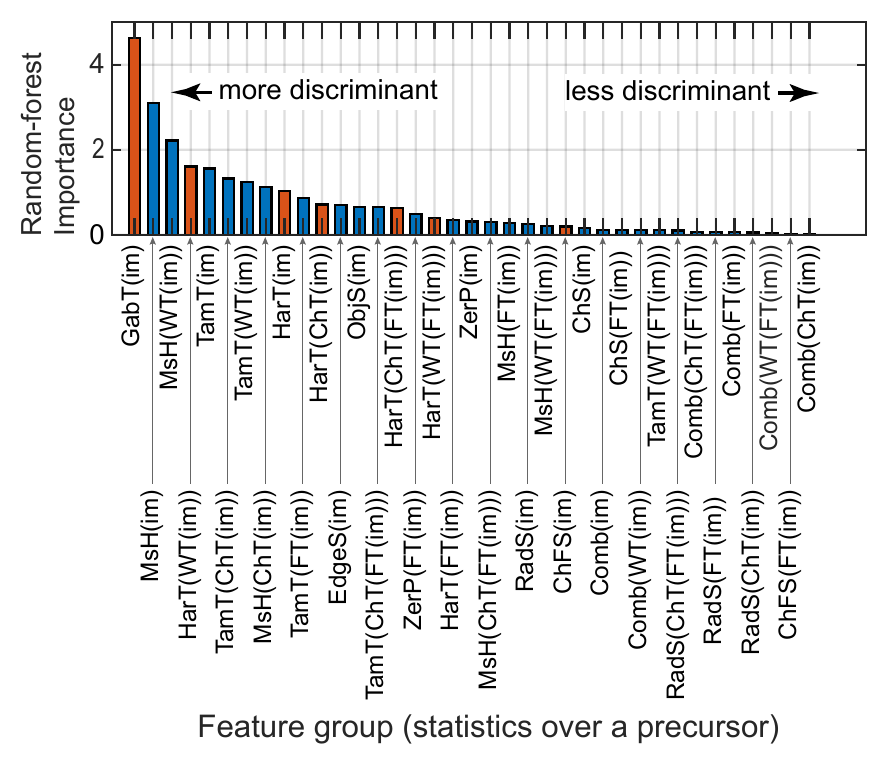}
			\caption{\textbf{Random forests using all the 1025 features} was trained and tested over 20\% of the dataset, and we retrieved the importance of each feature-group (see main text). Red bars highlight the feature groups displaying an execution time greater than \SI{20}{\milli\second}. The execution times are reported in Fig. \ref{fisher_scores_and_exec_time}a. The feature groups are described in Fig. \ref{fisher_scores_and_exec_time}c.} 
			\label{random_forest_full_dataset_result}
		\end{figure}
		
		The random forests offer a mechanism to assess the \textit{importance} of each feature in the decision  \citep{breiman_random_2001}. In a nutshell, it corresponds to the difference of the rate of misclassification of the "out-of-bag" samples (i.e. the labelled images not used for training a given tree because of the internal bootstrap mechanism), when randomly shuffling the values of a given feature. Hence, the importance of features is directly related to the performed classification, in contrast to Fisher's discriminant criterion used above. We summarised the feature importances as previously, by taking the average over their values within a group. We then averaged over the five forests generated in the \textit{k}-fold validation process (Fig. \ref{random_forest_full_dataset_result}). 
		
\begin{figure}[h!]
			\centering
			\includegraphics{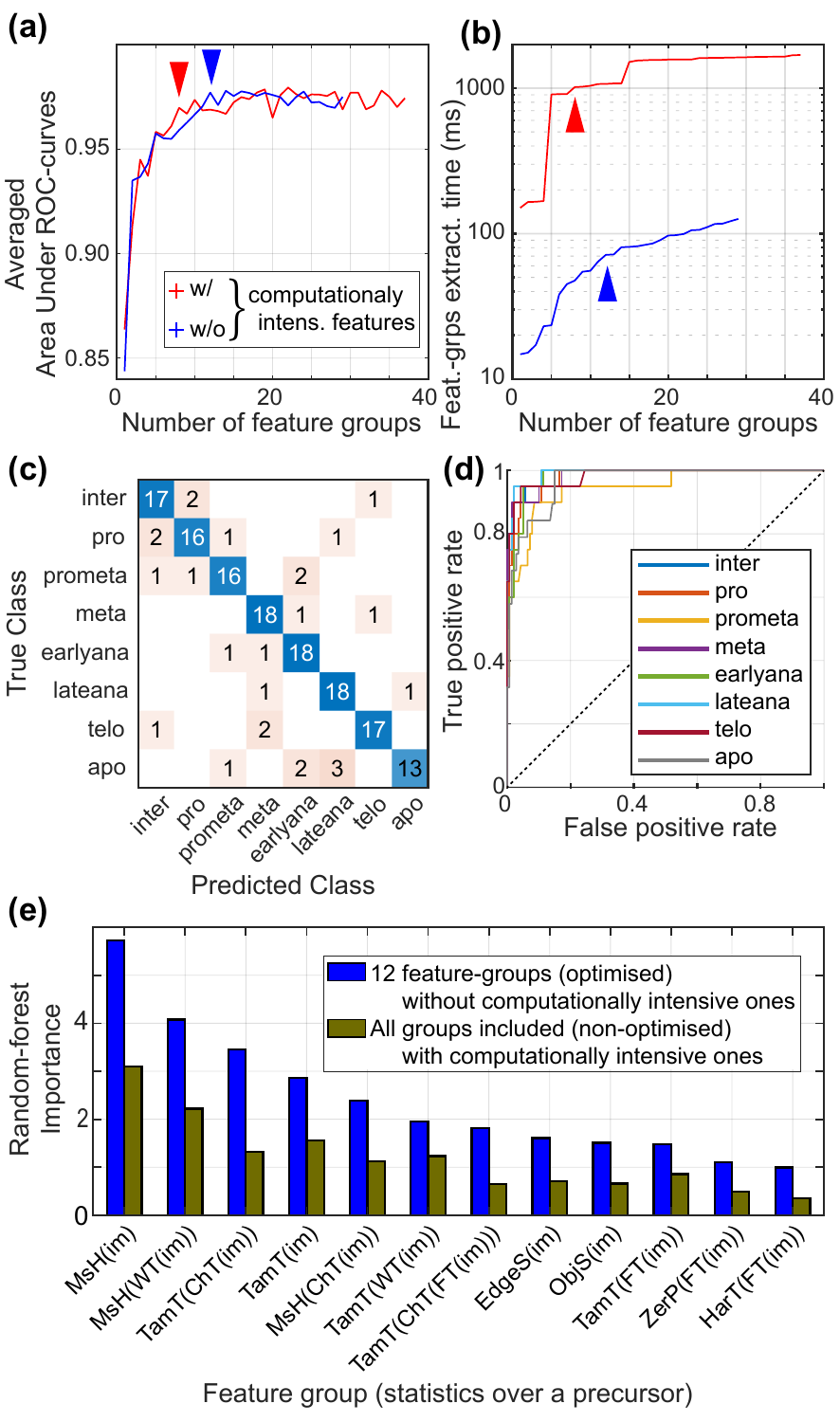}
			\caption{\textbf{Random-forests classification using a subset of feature-groups}. \textbf{(a)} Area Under Curve (AUC) averaged over the classes and \textbf{(b)}  execution time for extracting the feature-groups included in the classification, both versus the number of feature-groups used in classification, including (red curve) all available features or (blue curve) only feature groups with an execution time below \SI{20}{\milli\second} (not computationally intensive). Arrowheads of the corresponding colour depict their optimal number (see §\ref{RFoptim}). \textbf{(c)} Random forests importance (blue) in the twelve-groups case, optimal when  excluding computationally intensive feature-groups, and (brown) the all-feature-case (non optimised, reported Fig. \ref{random_forest_full_dataset_result} and \ref{random_forest_full_dataset_result_AUC_ROC}). We averaged over the 5-fold cross-validation and used the CellCognition dataset (see Methods §\ref{imDB}). \textbf{(d)} The confusion matrix and \textbf{(e)} the ROC curves averaged, using the 5-fold cross-validation in the twelve-feature-groups optimal case without computationally intensive feature-groups in the optimal case using the CellCognition dataset (see §\ref{imDB}). Class names are abbreviated after Fig. \ref{cell_classes}a.}
			\label{random_forest_reduced_dataset_result}
		\end{figure}

		\label{RFoptim}
		We aimed to perform a fast and precise classification, thus as for the Fisher's linear discriminant, we removed the least important feature groups and computed the random-forests importances again over the training vignettes and  averaged over 5-fold cross-validation. Unlike the case of Fisher's discriminant, the approach was iterative, requiring a re-training upon each change of the feature-group subset. The classification quality, measured by the mean AUC, decreased when using less than 8 groups (Fig.  \ref{random_forest_reduced_dataset_result}a, red curve). These 8 groups represented 147 features out of 1025. The global accuracy obtained with 8 groups was 75.5\% and the mean AUC 0.970, so very close to the results obtained with all features, suggesting that we could reduce the execution time by excluding features, without decreasing the classification quality (Fig. \ref{S3bis}).
	
			When it came to applying random forests to on-the-fly classification, we yet noticed that some computationally intensive feature-groups (red in Fig. \ref{random_forest_full_dataset_result}) displayed large importance like Gabor-on-raw-image and Haralick-on-wavelet-transform textures. On the ground of the trend obtained using Fisher's linear discriminant, we excluded the groups, which execution time was greater than \SI{20}{\milli\second}. We then iteratively removed the least-important features until it degraded the classification (Fig. \ref{random_forest_reduced_dataset_result}a, blue curve). It showed an optimum with 12 feature groups (264 features out of 1025). In that latter case, AUC read 0.977 and global accuracy 83.6\%, which was again very similar to the case using all 1025 features. We also obtained similar confusion matrix and the ROC curves (Fig. \ref{random_forest_reduced_dataset_result}cd) but the execution time was considerably reduced (divided by more than 50). This result validated the feasibility of our embedded classification by reducing the number of features and censoring the computationally intensive ones (Fig. \ref{random_forest_reduced_dataset_result}b).
	
		We then took a closer look at the feature importance when reducing the number of features, to get clues of this compensating mechanism. We compared the importance of the 12 feature-groups used in the optimised classification, with the importance of the same groups upon classifying over all the features (Fig. \ref{random_forest_reduced_dataset_result}e). We observed that the importance of these groups increased. It is suggestive of redundancy of the features, at least in their significance for the present classification if not in general. Notably, it was proposed that the random forests spread the importance among the redundant features. As expected, compensation of removed redundant features similar to what we observed was seen in other studies \citep{tuv_feature_2009, zhao_maximum_2019}. We here took advantage of this ability of random forests to ensure fast execution on an embedded system, compatible with real-time classification in an automated microscope.
	
			\label{RFbootstrap}
			We reckoned that these results represented one particular instance of database equilibration (see §\ref{imDB}). To test how general was our approach, we used bootstrap to randomly split data into balanced datasets, however without replacement (no duplicated image). We performed ten bootstrap iterations. Within each of them, we performed a 5-fold validation and repeated the optimisation process as described above, excluding computationally intensive feature-groups. On average, 12 feature-groups were the optimal balance between performance and accuracy (precisely \num{11.6 \pm 2.4}, mean $\pm$ standard deviation), as found previously, although variations of a few units were observed. We observed a \SI{79.6 \pm 2.4}{\percent} accurate classification lasting overall (feature extracting and vignette classification) \SI{68.7 \pm 3.5}{\milli\second}. Furthermore, the variations of classification accuracy and total execution time between bootstrap-iterations were reduced (Fig. \ref{random_forest_bootstrap}). However, the execution time varied mildly between each iteration of the bootstrap, in particular, because the selected feature-groups changed marginally in each bootstrap iteration. Indeed, we observed that 11 feature-groups are present in all these instances, while 1 is drawn in four other groups (black and blue text, Tab. \ref{random_forest_bootstrap:important_features_groups}). The low difference between bootstrap iterations showed the reproducibility of our method when different training subsets are used.
			 
	
		\begin{table}[h!]
				\centering
				\small
				\begin{tabular}{|cc|cc|}
					\toprule
					\textbf{Feature groups} & \textbf{\makecell{In \textit{n} boot- \\ strap iter.}} & \textbf{Feature groups} & \textbf{\makecell{In \textit{n} boot- \\ strap iter.}} \\
					\midrule
					\textit{EdgeS(im)} & 10 &
					MsH(ChT(im)) & 10 \\
					MsH(WT(im)) & 10 &
					MsH(im) & 10 \\
					ObjS(im) & 10 &
					\textit{TamT(ChT(FT(im)))} & 10 \\
					TamT(ChT(im)) & 10 &
					\textit{TamT(FT(im))} & 10 \\
					TamT(WT(im)) & 10 &
					\textit{TamT(im)} & 10 \\
					\textit{ZerP(FT(im))} & 10 & & \\
					\textcolor{blue}{HarT(FT(im))} & 5 &
				\textcolor{blue}{MsH(ChT(FT(im)))} & 3 \\
				\textcolor{blue}{MsH(FT(im))} & 1 &
				\textcolor{blue}{\textit{ZerP(im)}} & 1 \\
					\bottomrule
				\end{tabular}
			\caption{\textbf{Bootstrapping random forests optimal feature-groups-number classification} over the \textit{CellCognition} dataset. (black) 11/12 groups were always present in the 10 bootstrap iterations while (blue) the last group was taken among four other groups. The feature groups appearing only in the optimal cases using this dataset compared to \textit{mitocheck} one were italicised (Tab. \ref{random_forest_thomas_walter:important_features_groups}). The feature groups are described in Fig. \ref{fisher_scores_and_exec_time}c.}
				\label{random_forest_bootstrap:important_features_groups}
		\end{table}
		
\begin{figure}[h!p]
			\centering
			\includegraphics{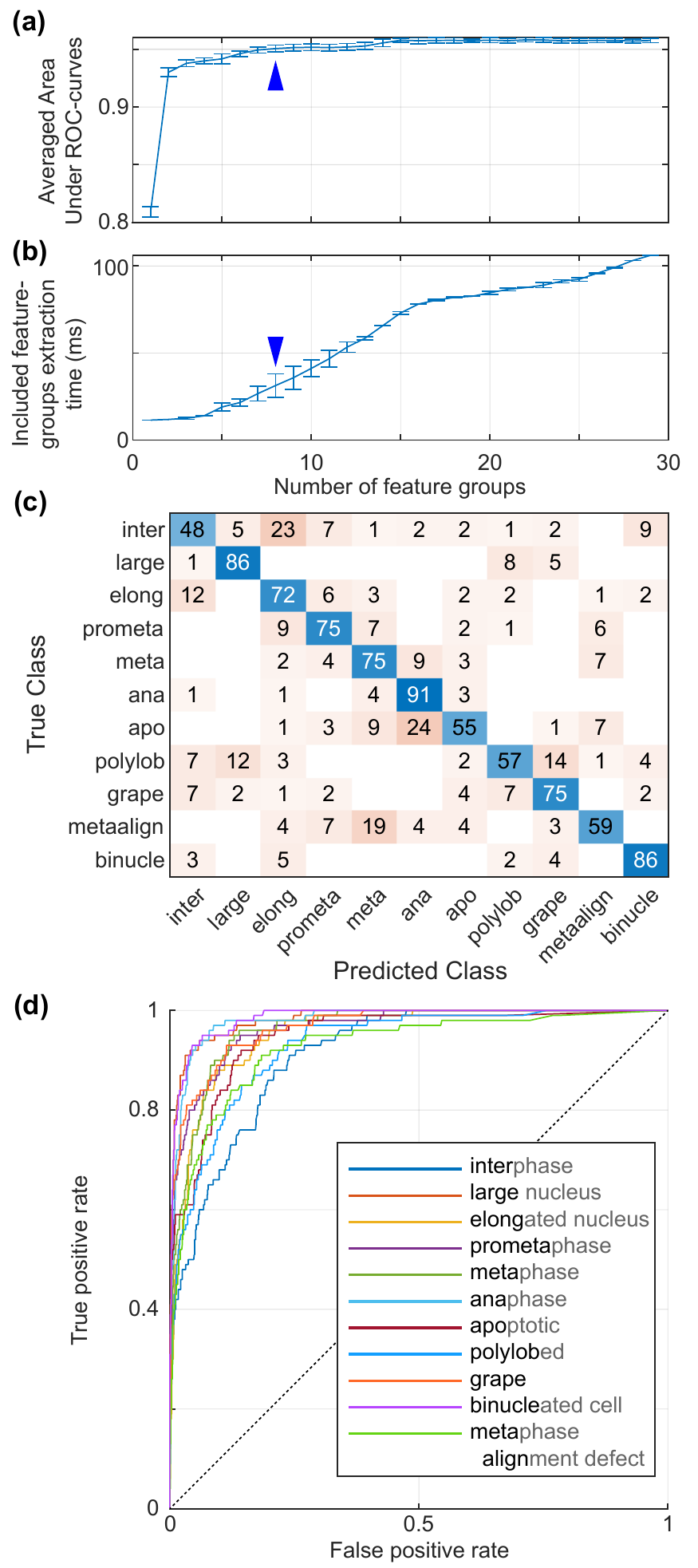}
			\caption{\textbf{Bootstrapping optimised random forests over \textit{mitocheck} dataset}. \textbf{(a)} The Area Under Curve (AUC) was averaged over the classes, and \textbf{(b)}  execution time for extracting the feature-groups included in classification was assessed (dependent of the selected feature-groups mildly variable between bootstrap iterations, see §\ref{RFbootstrap}). Both quantities are plotted versus the number of feature-groups used in classification and were computed in the 5-fold cross-validation repeats. This approach was repeated 10 times in the bootstrap approach, where the vignettes included in the balanced dataset were selected differently (see Methods §\ref{imDB}). We thus obtained the standard deviations reported by the error bars. Arrowheads depict the 8 feature groups optimal case. \textbf{(c)} The confusion matrix and \textbf{(d)} the ROC curves over the 5-fold cross-validation in a single bootstrap iteration.}
			\label{random_forest_thomas_walter}
		\end{figure}

		To further confirm this result, we repeated the approach using the second dataset \textit{mitocheck} (see §\ref{imDB}). In this case, images were classified between 11 classes, with 100 vignettes per class. We followed the same method as above and performed a \textit{k}-fold validation process ($k=5$) followed by ten bootstrap iterations, randomly spliting data into balanced datasets, however without replacement (no duplicated image). Eight feature groups, excluding the ones which execution time exceed \SI{20}{\milli\second}, were enough to achieve an optimal classification (Fig. \ref{random_forest_thomas_walter}ab). All classes were correctly recovered (Fig. \ref{random_forest_thomas_walter}cd). The feature-groups finally used in classification vary in the different instances of the bootstrap as with the CellCognition dataset without considerably impacting the execution time and the classification quality (Fig. \ref{random_forest_thomas_walter}). Interestingly, the selected feature-groups are mostly the same as with \textit{mitocheck} dataset (compare Tab. \ref{random_forest_thomas_walter:important_features_groups} and \ref{random_forest_thomas_walter:important_features_groups}). Overall, it confirmed the robustness of the above procedure used to embed image processing. Interestingly, most of the feature-groups were conserved, suggesting some possible generalisation.

		\begin{table}[h!]
				\centering
				\small
				\begin{tabular}{lcclcc|}
					\toprule
					\textbf{Feature groups} & \textbf{\makecell{In \textit{n} boot- \\ strap iter.}} & \textbf{Feature groups} & \textbf{\makecell{In \textit{n} boot- \\ strap iter.}} \\
					\midrule
					MsH(ChT(im)& 10 &
					MsH(im) & 10 \\
					MsH(WT(im)) & 10 &
					ObjS(im) & 10 \\
					TamT(ChT(im)) & 10 &
					TamT(WT(im)) & 10 \\
					\textcolor{blue}{\textit{MsH(WT(FT(im)))}} & 6 &
					\textcolor{blue}{HarT(FT(im))} & 4 \\
					\textcolor{blue}{MsH(ChT(FT(im)))} & 4 &
					\textcolor{blue}{MsH(FT(im))} & 4 \\
					\textcolor{blue}{\textit{TamT(WT(FT(im)))}} & 2 & & \\
					\bottomrule
				\end{tabular}
			\caption{\textbf{Bootstrapping random forests optimal feature-groups-number classification} over the \textit{mitocheck} dataset. (black) 6/8 groups were always present in the 10 bootstrap iterations while (blue) the two other groups were taken among five other groups. The feature groups appearing only in the optimal cases using this dataset compared to \textit{CellCognition} one, were italicised (Tab. \ref{random_forest_bootstrap:important_features_groups}). The feature groups are described in Fig. \ref{fisher_scores_and_exec_time}c.}
			\label{random_forest_thomas_walter:important_features_groups}
		\end{table}

		\label{time_RF}
		In the perspective of classifying vignettes on-the-fly, we had focused on the feature-extraction time by analogy to Fisher's linear discriminant, where this task took the vast majority of the execution time and where classification execution time was insignificant. Therefore, we had only embedded the feature extraction up to now to assess these times when using the random forests. We here reevaluated whether random forests might take significant time as it used decision trees. To do so, we ran the random forests classification on the embedded system using the RTrees module using the OpenCV library \citep{itseez2015opencv}. For the sake of simplicity, in a proof-of-concept perspective, we trained the algorithm using OpenCV on the embedded system. However, one could train on a general-purpose computer and embed only the classification. We then assessed the classification performance using 32 test vignettes (20\% of the whole \textit{CellCognition} dataset) in the optimal twelve-feature-groups case, excluding computationally intensive ones. With 300 trees, the execution time to classify these vignettes read \SI{89 \pm 20}{\micro\second} (mean $\pm$ standard deviation), extrapolated to \SI{27 \pm 6}{\milli\second} for a 300 cells picture. It should be compared to feature extraction over the same picture, lasting \SI{21.6}{\second}. Because feature extraction is performed independently on each vignette, this latter time could be scaled down by parallelising the features extraction since the NVIDIA Jetson AGX Xavier that we used here had 8 CPU cores. Finally, segmenting the image on one CPU core to extract the vignettes took a not noticeable time, about \SI{132 \pm 5}{\milli\second} (mean $\pm$ standard deviation) for the whole picture, in comparison to features extracting. In any cases, the classification itself took a lightweight time compared to the feature extraction.
		
		To conclude, we showed that using a non-linear method allowed us to find a much better time-performance compromise than the linear method, to both ensure fast and accurate classification. Therefore, we could envision using our feature-group optimised random forests together with the WND-CHARM features to enslave microscope driving to image classification.

	\subsection{Neural-network classification also benefits from feature-groups redundancy.}

		\begin{figure}[h!p]
			\centering
			
			\includegraphics{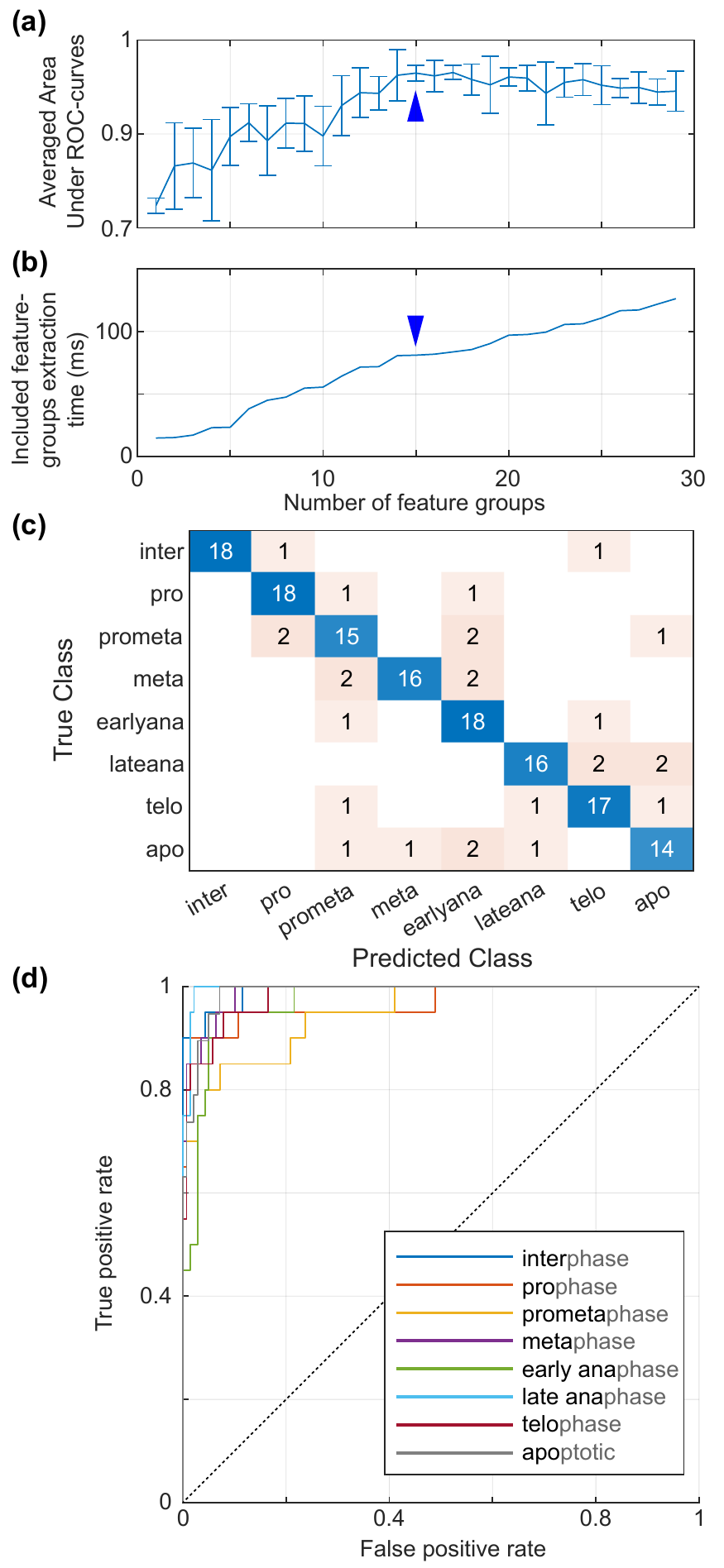}
			\caption{\textbf{Bootstrapping optimised neural network over \textit{CellCognition} dataset.} \textbf{(a)} The Area Under Curve (AUC) was averaged over the classes, and \textbf{(b)} execution time for extracting the feature-groups included in the classification was assessed. Both quantities are plotted versus the number of feature-groups used in classification and were computed in the 5-fold cross-validation repeats. This approach was repeated 20 times in the bootstrap approach, where the vignettes included in the balanced dataset were selected differently (see Methods §\ref{imDB}). We thus obtained the standard deviations reported by the error bars. Arrowheads depict the fiveteen-feature-groups optimal case. \textbf{(c)} The confusion matrix and \textbf{(d)} the ROC curves over the 5-fold cross-validation in a single bootstrap iteration. It is noteworthy that no error bar can be computed on execution time as the features are always ranked in the same order of importance (see main text §\ref{NN}).}
			\label{neural_net_results}
		\end{figure}

		\label{NN}
		Deep learning is the current paradigm in biological images analysis \citep{nagao20,moen19}. We wondered whether the proposed approach discarding highly discriminant features for the sake of rapidity keeping accuracy could be used in that context. We addressed this question in two steps firstly using a neural network as classifier and secondly extracting the features through the convolutional layers of a deep network classifier. Furthermore, fundamental research applications are more demanding about performances, requiring faster cycle time. Indeed, when it comes to studying mitotic events like metaphase-anaphase transition, the dynamics of the components are on the scale of the second or even the tenth of a second \citep{elting18}. To reach such fast processing, we could speed up the feature extraction through GPU-parallelisation, although it was out of the scope of the present paper. In such a context, the time spent in the classification itself became as well important. However, because of the high usage of conditional structures in such decision-tree-based methods, parallelising the random forests appeared difficult. We reckoned that we could use neural-network-based machine learning over the selected feature-groups. However, this method is more prone to overfitting \citep{tuv_feature_2009,bolon12}. In our case, this issue is worsened when the number of features is large, when they are non-independent, correlated or poorly informative as observed in our case. We, therefore, kept using the random forests to select the optimal feature groups, while we used the neural network in "production context" to perform classification.
				
		We trained a one-hidden-layer network with 64 neurons, using the gradient descent backpropagation algorithm with an adaptative learning rate starting from $0.01$, a momentum of $0.1$ and a mean squared error (MSE) loss function. To avoid over-fitting, an L2 regularisation parameter was added to the loss function with a $0.1$ ratio. These training parameters have been experimentally determined. The dataset was divided into three parts: training (70\%), validation (20\%) and test (10\%). The validation subset was used to stop training when the neural network started to overfit. As previously done, we used bootstrap to randomly split the whole data into a balanced dataset (see §\ref{imDB}), however without replacement (no duplicated image). We performed twenty bootstrap iterations. Within each of them, we used \textit{k}-fold cross-validation, with $k=10$. For each instance of the \textit{k}-fold process, the weights and biases of the networks were initialised to the same values. As previously, we tested the optimal number of feature-groups while excluding the most computationally intensive ones and assuming that group importances were ranked in the same order as in the case of random forests. The optimal classification was found with 15 feature groups and showed comparable accuracy with random forests (Fig. \ref{neural_net_results}cd), reading an AUC of 0.979 and global accuracy of \SI{83.0}{\percent}. However, the quality was more variable than with Random forests (Fig. \ref{neural_net_results}a) across the twenty bootstrap iterations. In a broader take, it validated the possibility to use a simple neural network with equal classification quality despite the small training set and a large number of features.

		We embedded our neural network using activation functions provided by the OpenCV library. As in the case of random forests, after proper training, we executed the classification of 32 test vignettes. The execution time read \SI{92 \pm 15}{\micro\second}. It could be extrapolated to \SI{28 \pm 4}{\milli\second} for an image containing 300 cells. It has to be compared to the time taken by the random forests to perform a similar task,  \SI{27 \pm 6}{\milli\second}. The neural network performed similarly to the random forests when run on CPU. However, it could be further accelerated in the specific case of the neural network using GPU parallelisation. These times remained small compared to the ones needed for feature extraction §(see \ref{time_RF}). Notably, neural networks used more features groups to perform classification with similar quality than random forests (15 versus 12), which can diminish neural networks interest for execution-time optimisation (Fig. \ref{neural_net_results}b). Conversely, Random forests were much slower than the neural network to be trained: training 300 decision trees using Random forests with 127 samples (\SI{80}{\percent} of the whole dataset) and 264 features (the 12 best feature-groups) took \SI{21}{\second} on Matlab using one CPU while training our neural network needed between \SIrange{1}{6}{\second}. The need for random forests to rank feature-groups by importance for each new category of images mitigated this advantage of the neural networks. Overall, the neural networks are more promising, but feature extraction will have to be parallelised to realise this pledge.
			
			\subsection{Features extracted through a convolutional neural network also show redundancy.}
\begin{figure}[h!p]
			\centering
			\includegraphics{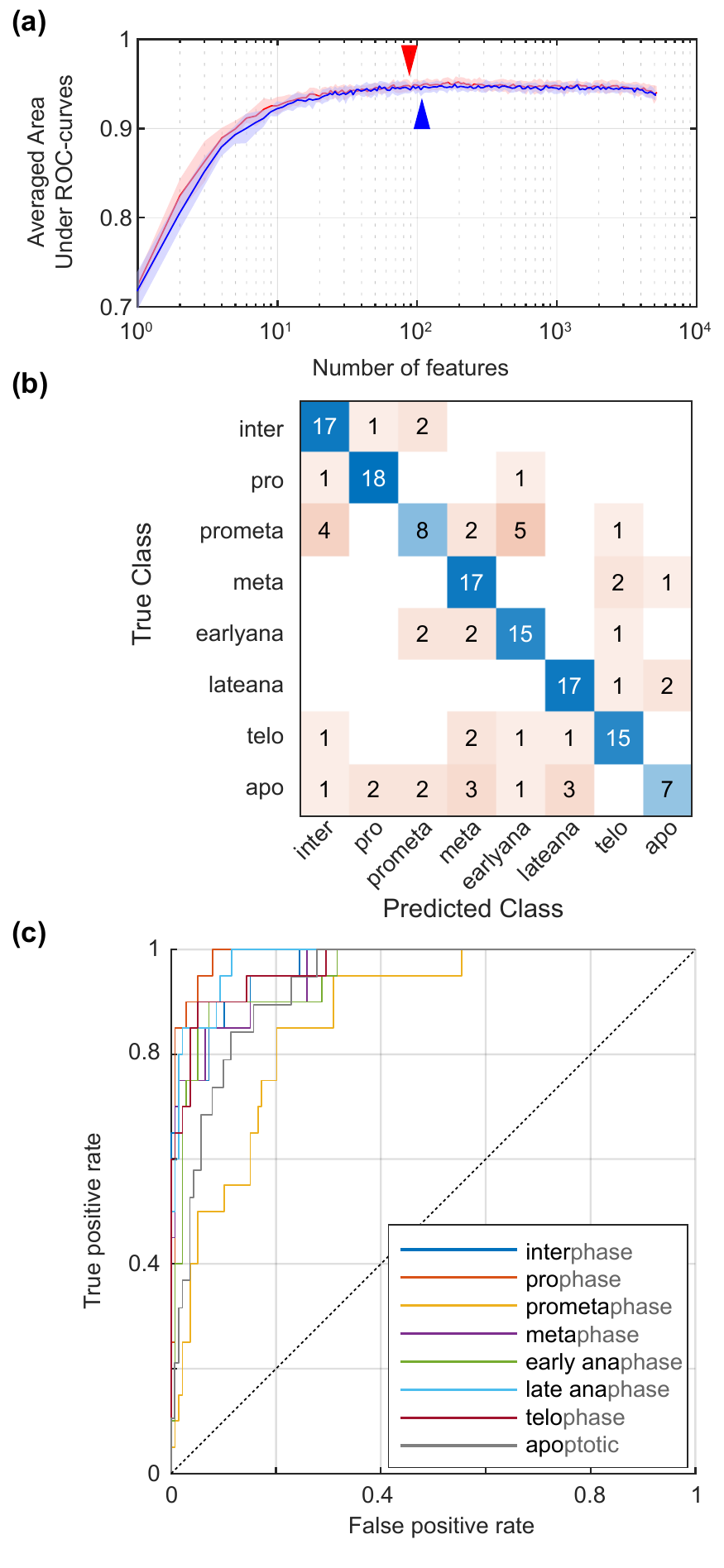}
			\caption{\textbf{Random forests classification extracting pseudo-features through a convolutional neural network and optimising the pseudo-feature number} over the \textit{CellCognition} dataset. \textbf{(a)} Area Under Curve (AUC) averaged over the classes versus the number of pseudo-features used in classification, including (red curve) all available pseudo-features or (blue curve) discarding the 100 most significant ones. Arrowheads of the corresponding colour depict their optimal number. \textbf{(b)} The confusion matrix and \textbf{(c)} the ROC curves averaged over the 5-fold cross-validation and ten bootstrap iterations, randomly splitting data into balanced datasets without duplicates (see §\ref{imDB}).}
			\label{CNN_results}
		\end{figure}

		We finally assessed whether the observed redundancy of biological images could be used to discard discriminant features in a deep neural network context. To do so, we built a simple convolutional neural network, including 3 convolutional layers separated by relu activation layers and trained it on the CellCognition images. We retrieved the outputs of the last layer before the fully-connected one and used them as pseudo-features. They are 5184, and we classified them using a 1000-trees random forests algorithm, to avoid over-fitting issues. We again performed 5-fold cross-validation followed by ten bootstrap iterations, randomly splitting data into balanced datasets, however without replacement (no duplicated image). We first included all the pseudo-features and iteratively reduced the number of feature by discarding the less important ones. We obtained an optimal classification with  \num{88 \pm 48} pseudo-features (mean $\pm$ standard deviation) (Fig. \ref{CNN_results}a, red curve). Fixing the number of pseudo-features to that number, we observed a larger variability of the pseudo-features included in the set among the bootstrap iterations. We might attribute it to observing single pseudo-features rather than groups; grouping would require a detailed analysis of the network out of the scope of this study. Consistently, among 275 pseudo-features appearing in one optimal set at least out of the ten bootstrap iterations, 18 are present in all sets and 71 in half of them at least. Overall, the optimal classification showed comparable accuracy with random forests, reading an averaged AUC of \num{0.948 \pm 0.006} and global accuracy of \SI{72 \pm 2}{\percent}.

			We then tested whether the compensating mechanism previously observed was applicable here. We thus suppressed the 100 most discriminant pseudo-features, i.e. reported as the most important by the random forests and selected in the optimal pseudo-feature set in at least 4/10 bootstrap iterations above. We repeated a similar analysis and obtained an optimal classification with  \num{108 \pm 124} pseudo-features (Fig. \ref{CNN_results}a, blue curve). Fixing the set to 108 pseudo-features, we observed an equivalent variability of the used pseudo-features as the case with all pseudo-features included: among 303 pseudo-features appearing in one optimal set at least out of the ten bootstrap iterations, 22 are present in all sets and 91 in half of them at least. The optimal classification displayed an accuracy similar to the case with all pseudo-features or with an optimal set among them not discarding important ones (Fig. \ref{CNN_results}a, compare red and blue curve tails and optimal pseudo-feature number marked by the arrowheads). In further details, we found an averaged AUC of \num{0.945 \pm 0.005} and global accuracy of \SI{71 \pm 2}{\percent}; the class-wise precisions were similar to the one obtained by classifying WND-CHARM features with random forests (Fig. \ref{CNN_results}bc). We concluded that pseudo-features based on deep-neural-networks convolutional layers were also redundant, allowing the most discriminant ones to be discarded. It proves that such a network could be pruned for the sake of computing time disregarding the importance of the nodes in classification.
	
	\section{Discussion and conclusion}
	
		In this study, we proposed a method to embed and execute cell-image classification in real-time as an essential module to create a smart microscope used for cell biology at large. In line with the reduced number of images available for training, a peculiar trait of our envisioned application, we used an existing general-purpose image feature extractor coupled with a machine learning algorithm. We analysed the contribution in the classifying decision of each feature, grouped by the image transforms from which they are computed. We took advantage of the machine learning algorithm that was able to report the feature importances. Doing so, we selected a subset of features best discriminating the various mitotic phases. Interestingly, censoring the most computationally intensive features did not degrade the classification upon re-training and selecting a new feature-subset. We could obtain excellent accuracy, suitable for the targeted application, by using a non-linear Machine Leaning method, combined with high execution performance on an embedded system to ensure analysis on-the-fly. In our example, we could classify about 14 cells per second into 8 phases of the cell cycle, with an accuracy greater than 80\% using Random forests classification. Using the almost the same subset of features, we can train a small neural network and reach similar performances benefiting of a classifier easy to embed and optimise on GPU. Importantly, this approach is transferable to deep learning network commonly used nowadays.
		
		Why suppressing the most discriminative features, for the sake of the execution time, did not degrade the classification accuracy? The various features, despite they belong to different groups and use a distinct strategy, are likely redundant. However, the quantity redundant to a censored feature is a non-linear combination of the available features as suggested by the better accuracy achieved when using a non-linear method. The replacing features are thus non-intuitive and likely not easily accessible by direct programming, outside of statistical modelling. Indeed, a large set of features as the one offered by WND-CHARM are expected to be redundant, and the use of decision trees appears well appropriate to decrease this redundancy \citep{tuv_feature_2009,bolon12}. Beyond this aspect, biological processes might also correlate some features independent mathematically-speaking. For instance, in the context of deep learning and larger image datasets, Nagao and co-authors found that additional markers on top of chromosomes did not improve the classification between the mitotic phases \citep{nagao20}. Indeed, the mitotic-phase changes involve numerous modifications of the sub-cellular structures, all under the control of the cell cycle regulation. It translates into various feature evolutions \citep{pollard02}. Along a similar line, measuring the mitotic spindle -- the essential structure tasked to dispatch the chromosomes to daughter cells correctly -- suggested that various features are correlated \citep{farhadifar15}. Similarly, we recently analysed the mitotic-spindle length and found that only three components, out of a principal component analysis, are enough to account for 95\% of inter-individual variability across more than 100 conditions obtained by involved protein depletion (Y. Le Cunff et al., data to be published). Overall, the variegated appearances of the sub-cellular structures as revealed by fluorescence microscopy are under the control of one or a few master regulators. Such a biological-originated correlation, modelled by our machine learning approach, further supports our strategy of reducing redundant features. While we investigated it on cell division, a similar situation likely happens in other cell-biology processes.

		The proposed methodology was developed keeping in mind that it should apply to small datasets, a constraint in application to biomedical science \citep{shaikhina15,kourou14,foster14}. Indeed, images are long to be produced and annotated. Furthermore, in the case of biological research, each experiment corresponds to a particular dataset: training with images from a distinct experiment (labelling other structures, e.g.) appears a poor option. As a result, only small datasets are available for training. This is a constraint shared with all experimental sciences and engineering, leading to reduce the number of degrees of freedom in the model, i.e. the number of used features and nodes in neural network \citep{feng19,pasupa16,shaikhina15,foster14}. The major risk is overtraining, leading the statistical model to learn details of the training set, failing to extract the general aspects, and \textit{in fine} causing low accuracy on real-data classification (testing). This is also why we opted here for the decision-tree forests and in particular, random forests, known to cope well with this issue at the first place\citep{breiman_random_2001,azar12}. Once this model is correctly trained, it helps to select features. Indeed, reducing the number of features, discarding the poorly-informative ones, not only improves the execution time but also limits the risk of overfitting in an approach similar to classical dropout technique used in deep learning \citep{srivastava14,borisov06}. In conclusion, our approach offers both a feature selection strategy enabling to decide the balance between execution time and accuracy directly, but also enables to use a neural network in a second time, when in the production set-up.

		We obtained the presented results using machine learning. We also showed that the removal of the most significant pseudo-features of a deep neural network, i.e. the nodes of the last layer before the fully connected one, does not preclude an accurate classification. On this ground, one can now envision using deep learning, in particular, pruning the networks as we know that an optimal number of features could be found \citep{molchanov16}. It will also benefit from the nowadays standard GPU acceleration of convolutional networks. The proposed method, by enabling accurate classification under the constraint of real-time execution, paves the way towards smart microscopy. This novel instrument, beyond making feasible experiments on rare and brief phenomena, will extend the HCS towards High Throughput Experimenting: beyond the bare observation of the sample, it will enable deeper imaging and in the future photo-perturbations. For example, this will enable to challenge the effect of drugs by investigating much more intimate processes of the cell. Finally and in a shorter term, medicine and biology are currently restricted to analyse data \textit{a posteriori}, requiring to acquire a huge amount of images to sort them afterwards because most of them are information-scarce. The smart microscopy promises a more parsimonious approach.


\section*{Acknowledgments}

We thank Drs. Hélène Bouvrais, Youssef El Habouz, Sébastien Huet, Yann Le Cunff and Sébastien Le Nours for discussions about the project. JP was supported by a Centre National de la Recherche Scientifique (CNRS) ATIP starting grant and La Ligue nationale contre le cancer. We acknowledge the support from Rennes Métropole and Région Bretagne through the program PME 2018-2019 under the reference roboscope, and from the national research agency (ANR-19-CE45-0011). MT and JP acknowledge France-BioImaging infrastructure supported by the French National Research Agency (ANR-10-INBS-04). This work was funded in part by the French government under management of Agence Nationale de la Recherche as part of the "Investissements d'avenir" program, reference ANR-19-P3IA-0001 (PRAIRIE 3IA Institute). MB fellowship was partially funded by the ANRT CIFRE program \#2017-1589. The University of Rennes 1  and Région Bretagne funded the fellowship of FS.

%
%

\bibliographystyle{model2-names}\biboptions{authoryear}
\bibliography{embedded_image_processing_MIA}

%


\clearpage
\section*{Supplemental figures}
\setcounter{figure}{0}
\renewcommand{\thefigure}{S\arabic{figure}}

		\begin{figure}[h]
			\centering
				\includegraphics{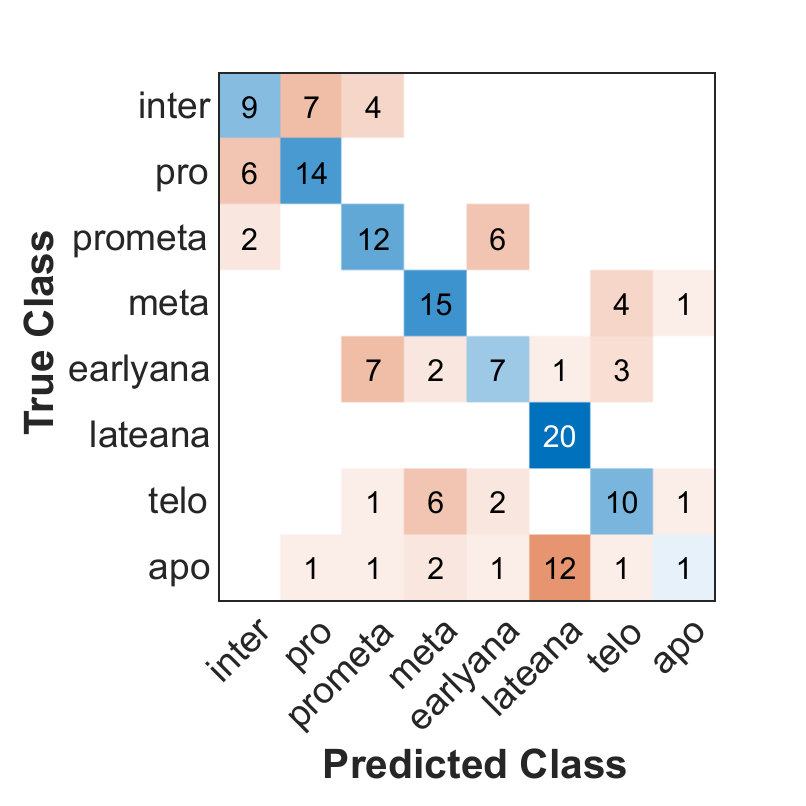}			
			\caption{\textbf{Classification using a single feature (Otsu-segmented-region area)} resulted in a poor confusion matrix. Class names are abbreviated after Fig. \ref{cell_classes}a. CellCognition dataset was used (see Methods §\ref{imDB}).}
			\label{confMatOtsu}
		\end{figure}

		\begin{figure}[h]
			\centering
				\includegraphics{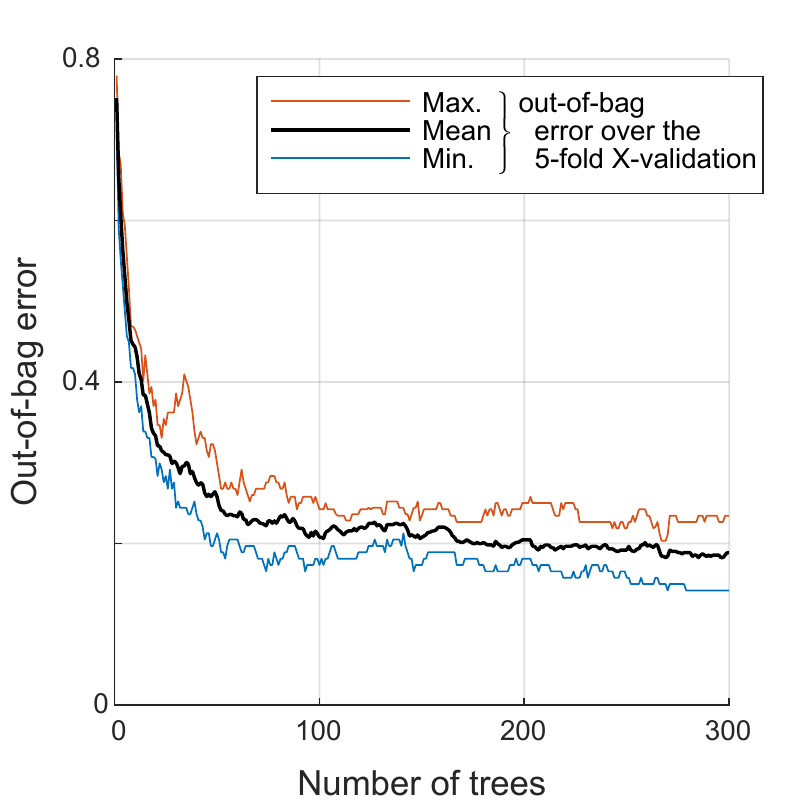}			
			\caption{\textbf{Selecting the number of trees in random forests classifier} by plotting the out-of-bag error versus the number of trees. The black, blue and red lines depict the average, minimum and maximum out-of-bag errors, respectively, over the 5-fold iterations of the cross-validation. CellCognition dataset was used (see Methods §\ref{imDB}).}
			\label{RFnbTree}
		\end{figure}

		\begin{figure}[h!]
			\centering
			
			\includegraphics{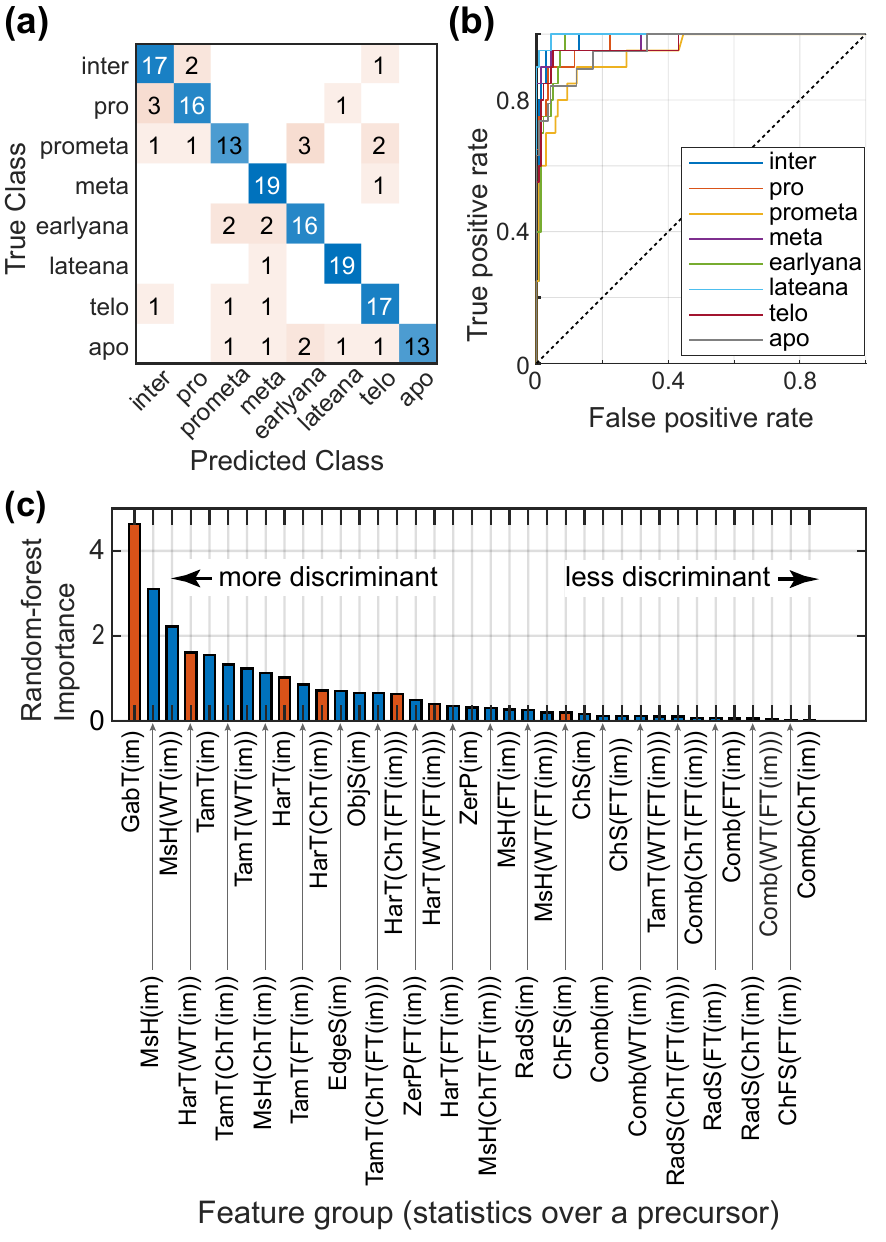}
					
			\caption{\textbf{Random forests using all the 1025 features} was trained and tested over 20\% of the dataset to get \textbf{(a)} the confusion matrix and \textbf{(b)} the ROC curves over the 5-fold cross-validation using the CellCognition dataset (see Methods §\ref{imDB}). Class names are abbreviated after Fig. \ref{cell_classes}a.} 
			\label{random_forest_full_dataset_result_AUC_ROC}
		\end{figure}

		\begin{figure}[h!]
			\centering
			
			\includegraphics{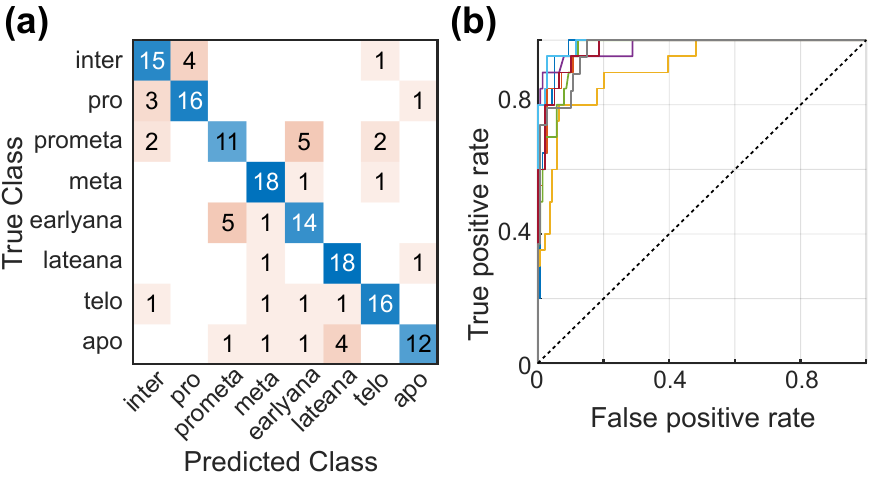}
					
			\caption{\textbf{Random forests with computationally intensive features} optimised by removing low importance feature groups. The algorithm was trained and tested over 20\% of the dataset to get \textbf{(a)} the confusion matrix and \textbf{(b)} the ROC curves over the 5-fold cross-validation using the CellCognition dataset (see Methods §\ref{imDB}). Class names are abbreviated after Fig. \ref{cell_classes}a.} 
			\label{S3bis}
		\end{figure}

			\begin{figure}[h]
			\centering
			\includegraphics{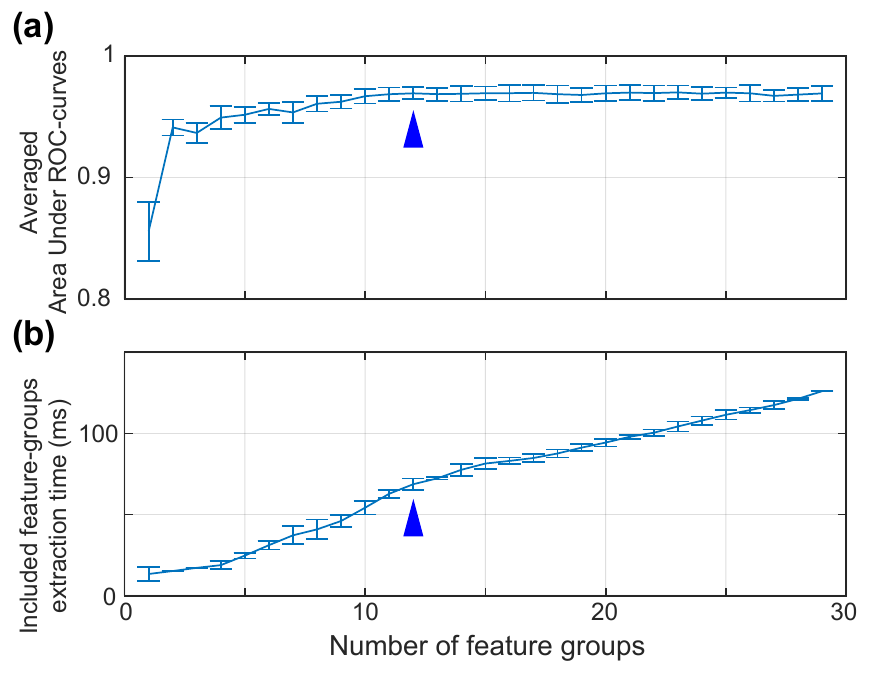}
			\caption{\textbf{Bootstrapping the random forests optimised with only non-computationally-intensive feature-groups}. \textbf{(a)} The Area Under Curve (AUC) was averaged over the classes, and \textbf{(b)} execution time for extracting the feature-groups included in the classification was assessed. Both quantities are plotted versus the number of feature-groups used in classification and  were computed in the 5-fold cross-validation repeats. This approach was repeated 10 times in the bootstrap approach, where the vignettes included in the balanced dataset were selected differently from the CellCognition (see Methods §\ref{imDB}). We thus obtained the standard deviations reported by the error bars. Fig. \ref{random_forest_reduced_dataset_result}ab report results in the same conditions for a single bootstrap iteration. Arrowheads depict the 12 feature groups optimal case.} 
			\label{random_forest_bootstrap}
		\end{figure}

\end{document}